# Beyond *Chandra* – the X-ray Surveyor

Martin C. Weisskopf[1a], Jessica Gaskin[a], Harvey Tananbaum[b], Alexey Vikhlinin[b]

[a]NASA/MSFC, ZP12, 320 Sparkman Drive, Huntsville, AL 35805; [b]Smithsonian Astrophysical Observatory, 60 Garden Street, Cambridge, MA 02138

## ABSTRACT

Over the past 16 years, NASA's Chandra X-ray Observatory has provided an unparalleled means for exploring the high energy universe with its half-arcsecond angular resolution. Chandra studies have deepened our understanding of galaxy clusters, active galactic nuclei, galaxies, supernova remnants, planets, and solar system objects addressing most, if not all, areas of current interest in astronomy and astrophysics. As we look beyond Chandra, it is clear that comparable or even better angular resolution with greatly increased photon throughput is essential to address even more demanding science questions, such as the formation and subsequent growth of black hole seeds at very high redshift; the emergence of the first galaxy groups; and details of feedback over a large range of scales from galaxies to galaxy clusters. Recently, NASA Marshall Space Flight Center, together with the Smithsonian Astrophysical Observatory, has initiated a concept study for such a mission now named the X-ray Surveyor. This concept study starts with a baseline payload consisting of a high resolution X-ray telescope and an instrument set which may include an X-ray calorimeter, a wide-field imager and a dispersive grating spectrometer and readout. The telescope would consist of highly nested thin shells, for which a number of technical approaches are currently under development, including adjustable X-ray optics, differential deposition, and modern polishing techniques applied to a variety of substrates. In many areas, the mission requirements would be no more stringent than those of Chandra, and the study takes advantage of similar studies for other large area missions carried out over the past two decades. Initial assessments indicate that such an X-ray mission is scientifically compelling, technically feasible, and worthy of a high prioritization by the next American National Academy of Sciences Decadal Survey for Astronomy and Astrophysics.

**Keywords:** X-ray Astronomy, X-ray optics, X-ray gratings, X-ray detectors

## 1. INTRODUCTION

The X-ray Surveyor is a candidate mission to bring to the attention of the USA National Academy of Science's 2020 Decadal Survey for Astronomy and Astrophysics. A NASA Headquarters Astrophysics Division white paper provided an initial list of missions drawn from the 2010 Decadal Survey and 2013 Astrophysics Roadmap that includes the X-ray Surveyor. The white paper also requested the three NASA Program Analysis Groups (PAGs) to coordinate community discussion to review and update the list of missions and instructed that PAG report(s) will be sent to the Astrophysics Subcommittee and then to the Astrophysics Division for selection of mission concepts to study. The process will result in appointments of Science and Technology Definition Teams and assignments of a lead NASA Center for each study. We represent a group of scientists that have some definite ideas as to what the X-ray Surveyor's capabilities could be.

## 2. THE INFORMAL MISSION CONCEPT TEAM

The Informal Mission Concept team is comprised of the lead authors of this paper together with the following with home institutions in parentheses: S. Bandler (NASA/Goddard Space flight Center - GSFC), M. Bautz (Massachusetts Institute of Technology - MIT), D. Burrows (Pennsylvania State University - PSU), A. Falcone (PSU), F. Harrison (California Institute of Technology), R. Heilmann (MIT), S. Heinz (University of Wisconsin), C.A. Kilbourne (GSFC), C. Kouveliotou (George Washington University), R. Kraft (Smithsonian Astrophysical Observatory - SAO), A. Kravtsov (University of Chicago), R. McEntaffer (University of Iowa), P. Natarajan (Yale University), S.L. O'Dell (NASA/Marshall Space Flight Center - MSFC), A. Ptak (GSFC), R. Petre (GSFC), B.D. Ramsey (MSFC), P. Reid (SAO), D. Schwartz (SAO), L. Townsley (PSU).

---

[1]martin.c.weisskopf@nasa.gov

## 3. BUILDING ON THE CHANDRA HERITAGE

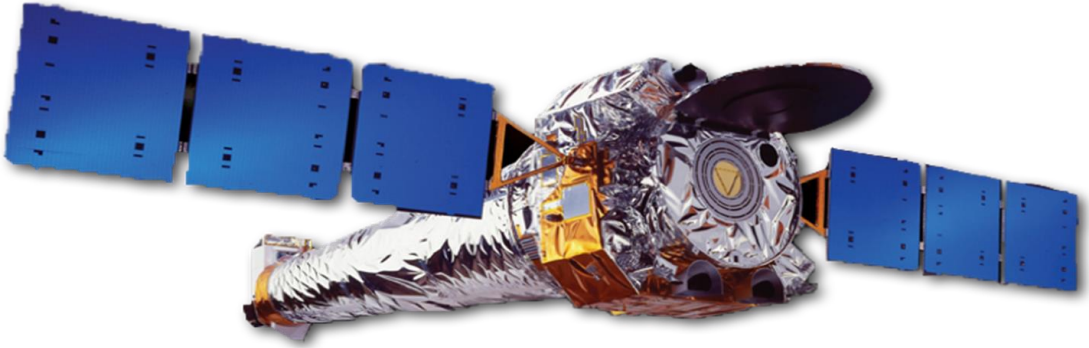

Figure 1. Artist's conception of the Chandra X-ray Observatory.

There can be no question that studies with Chandra have deepened our understanding of systems as diverse as galaxy clusters, active galaxies, normal and starburst galaxies, supernova remnants, normal stars, planets, and solar system objects.[1,2,3] The key to Chandra's success is the ½ arcsecond angular resolution. Notwithstanding paradigm-changing discoveries, it is also clear that many Chandra observations are photon-limited. Thus the baseline X-ray Surveyor concept is a successor to Chandra that has angular resolution at least comparable to Chandra, yet has much higher photon throughput. Maintaining a focal length approximately the same as Chandra's allows the new mission to limit spacecraft requirements to Chandra-like and result in a Chandra-like cost assuming that all necessary technologies will be at Technology Readiness Level 6 at the time the mission enters the development phase and certainly no later than the preliminary Design review. As discussed in the following, we feel that this is achievable.

## 4. THE OPTICS

The baseline optics concept utilizes the segmented optics approaches that were considered for a number of previous concept studies (Constellation-X, the International X-ray Observatory - IXO, etc) thus taking full advantage of what was accomplished. The process is summarized in cartoon form in Figure 2. Our X-ray Surveyor baseline design (Figure 3) involves 292 nested shells within a 3-m outer diameter resulting in an effective area at 1 keV of a factor of 30 greater than Chandra.

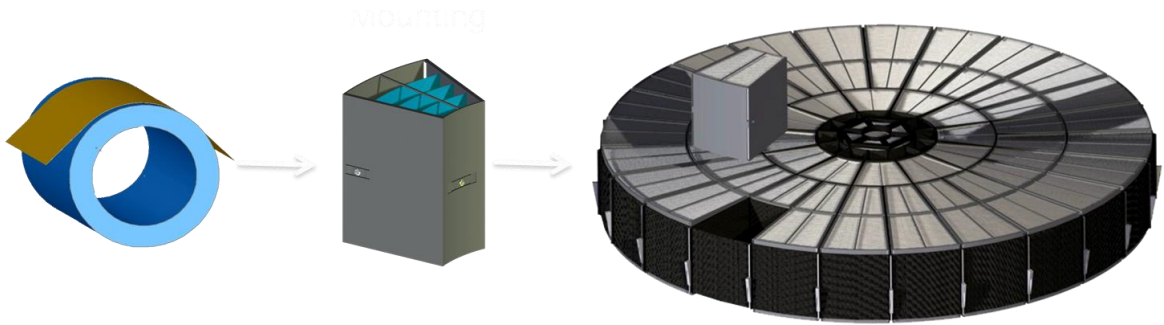

Figure 2. Illustration of the segmented optics approach. To the left is a representation of the formation of a reflecting element such as slumping thin glass on a precision mandrel. The elements (after possible additional processing) are then assembled and aligned into modules as illustrated in the central drawing. Modules are then assembled and aligned into a larger structure forming the telescope assembly.

The Surveyor telescope is based on a Wolter-Schwartzschild (W-S) design as opposed to the Chandra's Wolter-1. The W-S design increases the number of spatial resolution elements below a particular value for a given off-axis angle. The Surveyor optics are also much shorter than Chandra's which also flattens the field. Figure 4 shows the baseline Surveyor half-power diameter as a function of off-axis angle for the assumed ½-arcsecond on-axis resolution. Note the approximately 15-arcminute diameter field-of-view with one-arcsecond or better half-power diameter.

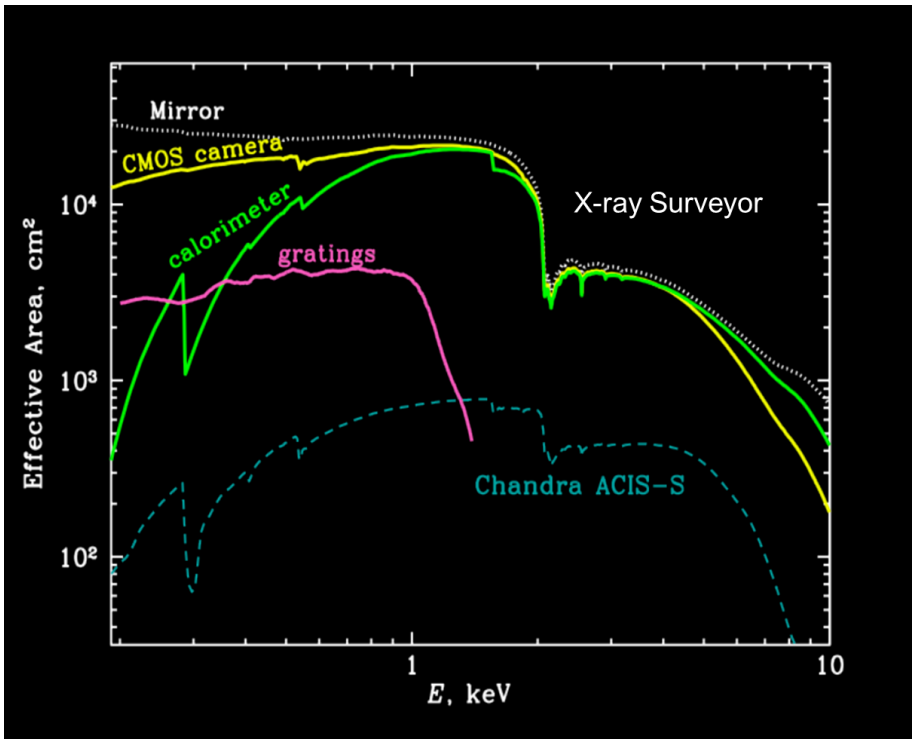

Figure 3. The baseline X-ray Surveyor effective area versus energy for different instruments and a comparison to Chandra.

Of course obtaining the sub-arcsecond reflecting elements will be a challenge. We are pursuing numerous approaches to this end since no current technology has yet demonstrated light-weight, sub-arcsecond optics. Currently, the two most promising approaches are differential deposition and adjustable optics. Slicing and forming thin polished silicon is also under preliminary investigation at the GSFC. The final approach may well be a combination of these.

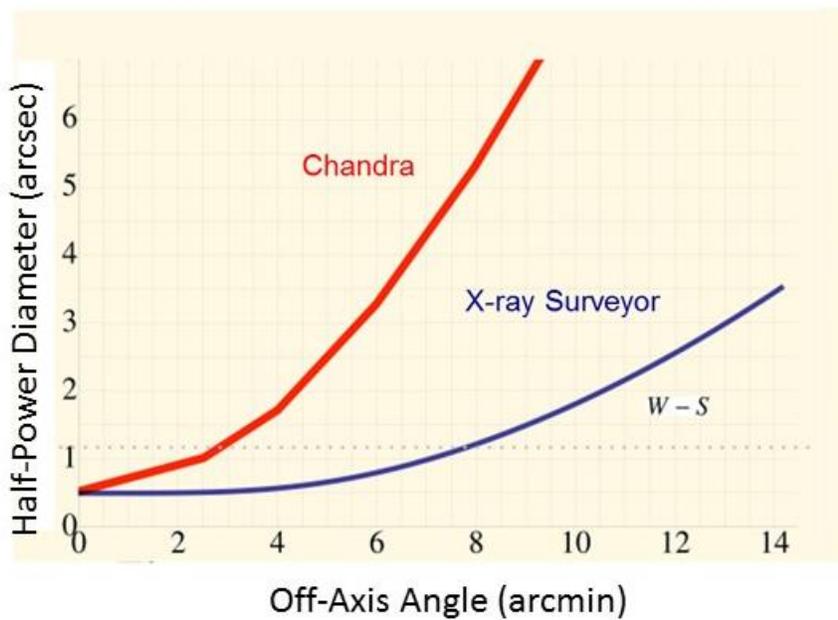

Figure 4. Half-power diameter versus off-axis angle (arcmin) for Chandra (top curve, red) and the X-ray Surveyor based on the Wolter-Schwartzschild design (bottom, blue).

**Differential Deposition**

Differential deposition, aka "filling in the valleys" is routinely used for thick synchrotron optics. There are at least two groups, one at NASA's MSFC and the other at Reflective X-ray Optics Corporation in New York, working with differential deposition and thin substrates including polished silicon. The process is illustrated in Figure 5. The group at MSFC in an early phase of its development program has corrected a segment of a full shell that originally exhibited a half-power-diameter of 7.1 arcseconds to 2.9 arcseconds after two deposition passes. The result was metrology limited. More details of the optics program at MSFC are discussed in paper 9510-2.

**Adjustable Optics**

Two approaches to adjustable X-ray optics are being pursued. These involve placing either a piezoelectric film (SAO/PSU) or a magneto-strictive film (Northwestern University) on the back of a thin reflecting surface. The former then applying and maintaining the appropriate voltage, the latter embedding the appropriate magnetic field. Work with magneto-strictive films is in its early stages.

Considerable progress has been made in the development of the piezoelectric adjustable optics, which through row-column addressing offer the capability to perform on-orbit adjustment of the telescope figure. Both a photo and an illustration of the concept are to be found in Figure 6. The method does not require a reaction structure. The approach has been shown to achieve micron-level corrections with <10V applied to 5–10 mm cells on segments 0.4-mm thick. Tests in the PSU University laboratory have >90% yield and ~5% uniformity on curved optical segments. Tests have shown that the response in the optical elements from individual piezoelectric cells is a good match to that expected from finite analysis modeling. Finally, stresses induced by the piezo deposition process itself can be compensated by the X-ray reflective coating.

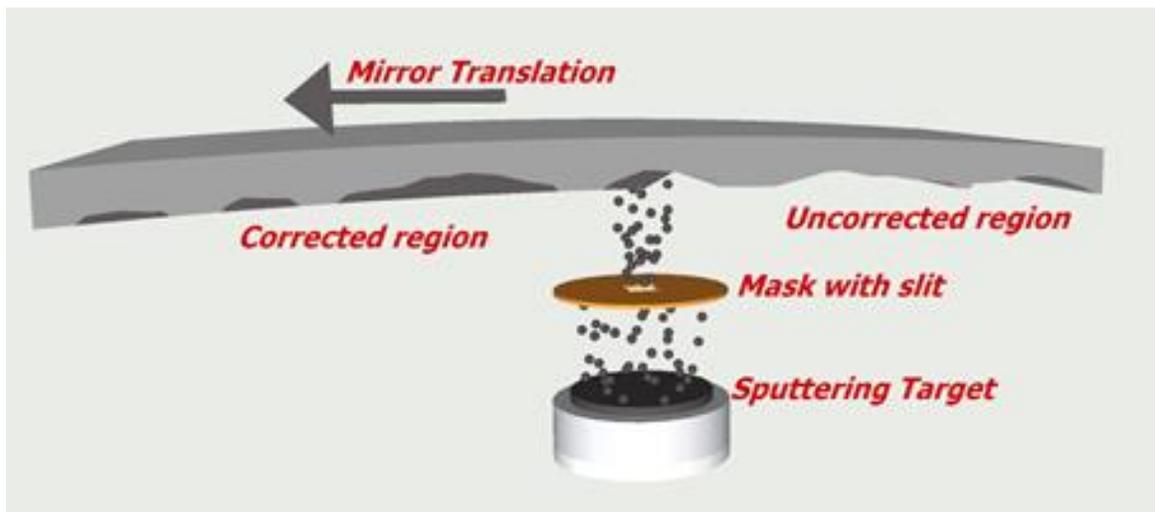

Figure 5. Illustration of the process of differential deposition. Varying the speed of the translation of the optic with respect to the sputtering target results in a corrected surface.

## 5. INSTRUMENTS

The Surveyor baseline concept makes use of what we envisage as the next-generation of X-ray instrumentation that exploits the new telescope's properties. These are: a microcalorimeter with a 5′×5′ field-of-view filled with 1″ pixels covering the energy range from 0.2 to 10 keV and providing better than 5 eV resolution, a CMOS active pixel sensor with 0.33″ pixels covering a 22′×22′ field of view and sensitive to X-rays in the energy range from 0.2–10 keV, and gratings that may be placed in the optical path that provide high-efficiency spectroscopy with resolving power of 5000 or more at the low energies where the calorimeter resolving power is lower.

As with the optics, none of these devices currently exist in a form that can meet all of the Surveyor requirements but there is significant activity and substantial progress towards the development of all of these instruments. [4,5,6,78]

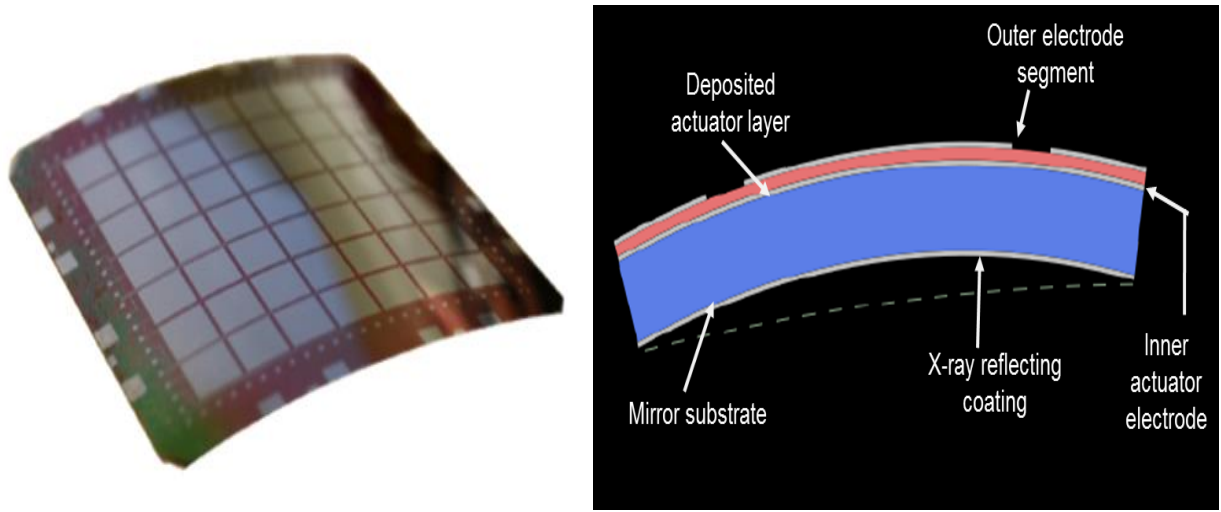

Figure 6. Left: Reflecting surface coated with an array of 7 × 7 1-cm pitch piezoelectric actuators. Right: Cartoon of the geometry.

## 6. CAPABILITY – AN EXAMPLE

The X-ray Surveyor represents a major leap forward for X-ray astronomy. As with Chandra, and indeed any major advance in observational capability, there will be many outstanding discoveries that we cannot yet predict. Nevertheless, the calculable advances are breath-taking. Consider e.g. a deep survey searching for the faintest and most distant X-ray emitting objects. The current most sensitive flux limit of ~5×10$^{-18}$ ergs/s/cm$^2$ (0.5-2 keV) is based on the 4 Msec exposure in the Chandra Deep Field South (which is being extended to 7 Msec at this time). Over the central 3' × 3' field of view where this flux limit could be reached, 69 sources were detected comprised of 32 galaxies and 37 active galactic nuclei (AGN). This should be compared to what may be accomplished with the X-ray Surveyor which, exploiting the sub-arcscond 15´ × 15´ field would, in the same 4 Msec integration time, detect 12,830 sources[9] comprised of 11060 galaxies and 1765 AGN and at a flux limit a factor of 35 lower than Chandra's. Of course the angular resolution avoids source confusion.

## 7. ACKNOWLEDGEMENTS

We wish to acknowledge detailed contributions from R. Hopkins, A. Schnell and the MSFC Advanced concepts office Team, P. Reid, R. Kraft, B. Ramsey, K. Kilaru, D. Burrows, A. Falcone, S. Bandler, C. Clairborne, R. McEntaffer and R. Heilmann.